\documentclass[aps,twocolumn,preprintnumbers,amsmath,amssymb,superscriptaddress,floatfix,nofootinbib]{revtex4}
\usepackage{graphicx,color,dcolumn,booktabs,bm}
\usepackage{longtable,lscape}
\usepackage{txfonts}
\usepackage{overpic}
\usepackage{epsfig}
\usepackage{amssymb}
\usepackage{rotating}
\usepackage{epstopdf}
\usepackage{appendix}
\usepackage{indentfirst}
\usepackage{feynmf}   %{feynmp}
\usepackage{slashed}  %for Feynman symbols
\usepackage{cases}
\usepackage{color}
\usepackage{multirow}
\usepackage{graphicx,color,dcolumn,booktabs,bm}
\usepackage{cases}
\usepackage{array}

\graphicspath{{Figures/}} %
\usepackage[colorlinks=true,citecolor=blue,anchorcolor=red,menucolor=red, linkcolor=red,filecolor=red,runcolor=red,urlcolor=blue,frenchlinks=true ]{hyperref}
\newcommand{\vsig}{\mbox{\boldmath$\sigma$\unboldmath}}

\newcommand{\vlab}{\mbox{\boldmath$\lambda$\unboldmath}}

\begin{document}
\title{Fully-heavy hexaquarks in a constituent quark model}

\author{Qi-Fang L\"u \footnote{Corresponding author} } \email{lvqifang@hunnu.edu.cn} %
\affiliation{ Department of Physics, Hunan Normal University, and Key Laboratory of Low-Dimensional Quantum Structures and Quantum Control of Ministry of Education, Changsha 410081, China }

\affiliation{Synergetic Innovation Center for Quantum Effects and 
Applications (SICQEA), Hunan Normal University, Changsha 410081, China}

\affiliation{Research Center for Nuclear Physics (RCNP), Ibaraki, 
Osaka 567-0047, Japan}

\author{Dian-Yong Chen} \email{chendy@seu.edu.cn} %
\affiliation {School of Physics, Southeast University, 
Nanjing 210094, China}
\affiliation{Lanzhou Center for Theoretical Physics, Lanzhou University, Lanzhou 730000, China}

\author{Yu-Bing Dong} \email{dongyb@ihep.ac.cn} %
\affiliation{Institute of High Energy Physics, Chinese Academy 
of Sciences, Beijing 100049, China}

%\affiliation{Theoretical Physics Center for Science Facilities (TPCSF), CAS, Beijing 100049, China}

\affiliation{School of Physical Sciences, University of Chinese Academy of Sciences, Beijing 101408, China}

\begin{abstract}
In this work, we systematically investigate the mass spectra of 
fully-heavy hexaquarks within a constituent quark model by 
including the color Coulomb potential, linear confining potential, 
and spin-spin interactions. Our results show that all of the 
fully-heavy hexaquarks lie above the corresponding 
baryon-baryon thresholds, and thus no stable compact one exists. 
These states may subsist as resonances and decay into two 
fully-heavy baryons easily through the fall-apart mechanism, 
which can be searched in future experiments.
\end{abstract}

\maketitle

\section{Introduction}
\label{sec:intro}

\quad Theoretical and experimental studies on the exotic states 
beyond the 
conventional hadrons play essential roles in understanding the 
nonperturbative properties of quantum chromodynamics (QCD). In the past 
two decades, researchers have witnessed lots of new hadrons observed by 
large-scale scientific facilities, and many of them are difficult to 
interpreted as the conventional hardons. With this remarkably 
experimental achievement, theorists also payed plenty of attentions 
and efforts to study the inner structures of these exotica and 
provided highly valuable information for experimental searches. More 
details of these discoveries and recent progresses can be found in the 
review articles~\cite{Hosaka:2016pey,Chen:2016qju,Lebed:2016hpi,
Esposito:2016noz,Dong:2017gaw,Ali:2017jda,Guo:2017jvc,Olsen:2017bmm,
Karliner:2017qhf,Liu:2019zoy,Brambilla:2019esw,Richard:2019cmi,
Barabanov:2020jvn,Chen:2022asf}.\\ 

\quad Those new exotica are usually interpreted as compact multi-quark 
states, loosely bound molecules, kinematic effects, or even the 
conventional states. Actually, people often have difficulties 
distinguishing the various explanations if they lie in the similar 
energy region and have same quantum numbers. Fortunately, the 
fully-heavy exotics are particularly interesting and less 
troublesome, where they are far away form the scope of conventional 
hadrons. Also, without the light meson exchanges, the loosely bound 
molecular configurations seem to be disfavored. Moreover, the
relativistic effects should be small enough owing to the absence 
of light quark, and the nonrelativistic quark model can describe 
these fully-heavy systems pretty well. Thus, it is a good place 
to hunt for the genuine compact exotic states.\\

\quad Since the observations of $X(6900)$ by the LHCb Collaboration 
in 2020~\cite{Aaij:2020fnh}, the fully-heavy exotics have attracted 
lots of interests in recent 
years~\cite{liu:2020eha,Wang:2020ols,Jin:2020jfc,Lu:2020cns,
Becchi:2020uvq,Chen:2020xwe,Wang:2020gmd,Sonnenschein:2020nwn,
Giron:2020wpx,Maiani:2020pur,Richard:2020hdw,Chao:2020dml,Wang:2020wrp,
Dong:2020nwy,Ma:2020kwb,Feng:2020riv,Zhao:2020nwy,Gordillo:2020sgc,
Weng:2020jao,Guo:2020pvt,Feng:2020qee,Zhang:2020vpz,Gong:2020bmg,
Dong:2020hxe,Wan:2020fsk,Liu:2020tqy,Zhao:2020cfi}. Also, more 
structures in the $J/\psi-J/\psi$ and  $J/\psi-\psi(2S)$ invariant mass 
spectrum very recently~\cite{CMS,ATLAS}. These structures are good 
candidates for the fully-heavy tetraquarks, which have been suggested in 
the literature. Also, the masses of the ground states for fully-heavy 
pentaquarks are investigated within the framework of quark 
model~\cite{An:2020jix,An:2022fvs}.\\

\quad Unlike the thriving scene of the tetraquarks and pentaquarks,  
the experimental and theoretical progresses on dibaryons seem 
relatively few. The deuteron is certainly a well established six-quark 
state, which is interpreted as a proton-neutron molecular state. 
Another six-quark state is the $d^*(2380)$ observed by the WASA-at-COSY 
Collaboration~\cite{Bashkanov:2008ih,WASA-at-COSY:2011bjg}, and are 
investigated by several theoretical 
groups~\cite{Dyson:1964xwa,Yuan:1999pg,Bashkanov:2013cla,Gal:2013dca,
Gal:2014zia,Huang:2013nba,Park:2015nha,Dai:2005kt,Huang:2014kja,
Dong:2015cxa,Chen:2014vha,Huang:2015nja,Dong:2016rva,Lu:2017uey}. 
Moreover, other possible structures are also claimed experimentally, 
such as the isotensor $\Delta N$ dibaryon resonance 
$D_{21}$~\cite{WASA-at-COSY:2018zlh,WASA-at-COSY:2018poi} and 
proton–$\Omega$ correlation in heavy-ion collisions~\cite{STAR:2018uho}, 
and the readers interested in light hexaquarks is referred to the recent 
review~\cite{Clement:2017vnl,Clement:2020mab}. Owing to the complexities 
of light few-body systems, the nature of $d^*(2380)$ is a long-standing 
problem, which seems intractable and will stay in dispute in the near 
future. A more realistic and intelligent way to looking for the 
hexaquarks is turning to the heavy systems, where the structures may be 
picked out more easily by experiments.\\

\quad There have been several studies on the fully-heavy 
hexaquarks in the 
literature, which mainly concentrate on the $\Omega_{ccc}\Omega_{ccc}$, 
$\Omega_{ccc}\Omega_{bbb}$, and  $\Omega_{bbb}\Omega_{bbb}$ systems. 
Within the lattice QCD method, the authors calculated the mass of 
various dibaryons  and pointed out that the $\Omega_{ccb}\Omega_{cbb}$, 
$\Omega_{ccc}\Omega_{ccc}$, and $\Omega_{bbb}\Omega_{bbb}$ 
dibaryons and should be below their respective two-baryon 
thresholds~\cite{Junnarkar:2019equ,Lyu:2021qsh,Mathur:2022nez}.  Also, 
the one-boson-exchange model and QCD sum rule approach suggested that 
there should exist weakly bound ones~\cite{Liu:2021pdu,Wang:2022jvk}. 
However, several studies within quark models disfavored the bound 
hexaquarks~\cite{Richard:2020zxb,Alcaraz-Pelegrina:2022fsi,Weng:2022ohh} 
except the work of Ref.~\cite{Huang:2020bmb}. In short, the previous 
works between quark models and quantum field theory approaches 
show quite different conclusions. Thus, it is interesting for us 
to investigate this controversy and find out possible reasons. 
Moreover, a systematical and solid study in a constituent quark 
model with the same potential and parameters as conventional 
hadrons are essential to ensure the validity of our exploration.\\

\quad In this work, we adopt a nonrelativistic constituent quark 
model to investigate the mass spectra of fully-heavy hexaquarks 
within compact configurations. This framework composed of 
potential and parameters have been widely employed to study the 
conventional and tetraquark 
states~\cite{liu:2020eha,Deng:2016stx,Liu:2019zuc,Liu:2019vtx,
Liu:2021rtn}, and have been proven effective enough for 
these fully-heavy systems. In the fully-heavy hexaquarks, the 
relativistic effects should be extremely small and no light meson 
exchange needs considering, which assures the reliability of results 
obtained by the nonrelativistic constituent quark model including 
Coulomb potential, linear confining potential, and spin-spin 
interactions. We find that all of the fully-heavy hexaquarks lie 
above the corresponding baryon-baryon thresholds, and thus no 
stable compact one exists. We hope our results can provide valuable 
information for future experimental and theoretical studies.    \\

\quad This article is organized as follows. In Section~\ref{model}, 
we introduce the framework of the nonrelativistic constituent quark 
model for hexaquarks. The results and discussions for the mass 
spectra of fully-heavy heavy hexaquarks are given in 
Section~\ref{results}. The last section is a summary.

\section{Framework} \label{model}

\quad In the nonrelativistic constituent quark model, 
the Hamiltonian of the fully-heavy hexaquark systems can be expressed as
\begin{equation}\label{Hamiltonian}
	H=\Bigg (\sum_{i=1}^6 m_i+T_i \Bigg )-T_G+\sum_{i<j}V_{ij}(r_{ij}),
\end{equation}
where $m_i$ and $T_i$ are the constituent quark mass and kinetic energy 
of the $i$th quark, respectively; $T_G$ stands for the center-of-mass 
kinetic energy of the hexaquark system; $V_{ij}(r_{ij})$ corresponds to 
the potential between the $i$-th and $j$-th quark, which includes 
short-range one-gluon-exchange interaction and long-range linear 
confinement. The explicit formula of $V_{ij}(r_{ij})$ can be written 
as~\cite{Liu:2019zuc} 
\begin{equation}\label{vij}
	V_{ij}(r_{ij})=V_{ij}^{OGE}(r_{ij})+V_{ij}^{Conf}(r_{ij})  \ ,
\end{equation}
with
\begin{equation}\label{voge}
	V^{OGE}_{ij}=\frac{\alpha_{ij}}{4}({\vlab}_i\cdot{\vlab}_j)\left\{\frac{1}{r_{ij}}-\frac{\pi}{2}\cdot\frac{\sigma^3_{ij}e^{-\sigma^2_{ij}r_{ij}^2}}{\pi^{3/2}}\cdot\frac{4}{3m_im_j}({\vsig}_i\cdot{\vsig}_j)\right\},
\end{equation}
and
\begin{equation}\label{vconf}
	V^{Conf}_{ij}(r_{ij})=-\frac{3}{16}({\vlab}_i\cdot{\vlab}_j)\cdot b r_{ij}.
\end{equation}
The relevant parameters are shown in Table~\ref{parameters}. With these
parameters, the mass spectra of fully-heavy mesons, baryons, and 
tetraquarks have been described well in previous 
works~\cite{liu:2020eha,Deng:2016stx,Liu:2019zuc,Liu:2019vtx,
Liu:2021rtn}. Also, some calculated masses of fully-heavy conventional 
hadrons are listed in Table~\ref{masses} for reference. Thus, it is 
suitable to adopt the same framework to investigate the fully-heavy 
hexaquark systems. \\

\begin{table}[htp]
	\begin{center}
		\caption{\label{parameters} Relevant parameters in the constituent quark model.}
		\begin{tabular}{cccccccccccc}\hline\hline
			~~~~~~&  $m_c$~(GeV)           &~~~~~~~~~~~~~~~~~~~~~~~~~~~~~~~~~~~~~~1.483~~~~~~~~~~~~~\\
			~~~~~~&  $m_b$~(GeV)           &~~~~~~~~~~~~~~~~~~~~~~~~~~~~~~~~~~~~~~4.852~~~~~~~~~~~~~\\
			~~~~~~&  ${\alpha_{cc}}$        &~~~~~~~~~~~~~~~~~~~~~~~~~~~~~~~~~~~~~~0.5461~~~~~~~~~~~~~\\
			~~~~~~&  ${\alpha_{bb}}$        &~~~~~~~~~~~~~~~~~~~~~~~~~~~~~~~~~~~~~~0.4311~~~~~~~~~~~~~\\
			~~~~~~&  ${\alpha_{bc}}$        &~~~~~~~~~~~~~~~~~~~~~~~~~~~~~~~~~~~~~~0.5021~~~~~~~~~~~~~\\
			~~~~~~&  ${\sigma_{cc}}$~(GeV) &~~~~~~~~~~~~~~~~~~~~~~~~~~~~~~~~~~~~~~1.1384~~~~~~~~~~~~~\\
			~~~~~~&  ${\sigma_{bb}}$~(GeV) &~~~~~~~~~~~~~~~~~~~~~~~~~~~~~~~~~~~~~~2.3200 ~~~~~~~~~~~~~\\
			~~~~~~&  ${\sigma_{bc}}$~(GeV) &~~~~~~~~~~~~~~~~~~~~~~~~~~~~~~~~~~~~~~1.3000 ~~~~~~~~~~~~~\\
			~~~~~~&  ${b}$ ~(GeV $^2$)    &~~~~~~~~~~~~~~~~~~~~~~~~~~~~~~~~~~~~~~0.1425~~~~~~~~~~~~~\\
			\hline\hline
		\end{tabular}
	\end{center}
\end{table}

\begin{table*}[!htbp]
	\begin{center}
		\caption{ \label{masses} Masses of fully-heavy conventional hadrons. Experimental date are taken from PDG~\cite{ParticleDataGroup:2020ssz}. The units are in MeV.}
		\begin{tabular*}{18cm}{@{\extracolsep{\fill}}*{13}{p{1.25cm}<{\centering}}}\hline\hline
			State   & $\eta_c$  & $J/\psi$& $B_c$  & $B_c^*$ &  $\eta_b$  & $\Upsilon$ & $\Omega_{ccc}$  &  $\Omega_{ccb}$  &  $\Omega_{ccb}^*$  &   $\Omega_{cbb}$  &  $\Omega_{cbb}^*$  &   $\Omega_{bbb}$      \\ \hline
			Mass  &2983  & 3097  & 6271  & 6326  &    9390     &      9460         & 4828                & 8047               & 8070        & 11248              & 11272               & 14432\\
			Experiments     & 2984  &  3097 & 6271  & $\cdots$  & 9399       &     9460           & $\cdots$                & $\cdots$         & $\cdots$               & $\cdots$               & $\cdots$  & $\cdots$\\
			\hline\hline
		\end{tabular*}
	\end{center}
\end{table*}

\quad To solve this Hamiltonian, we also need to construct the fully 
antisymmetric color-spin-flavor-orbital wave functions according to the 
Pauli exclusion principle. Since the charm and bottom quarks are not 
identical particles, the flavor parts of these systems are always 
trivial, that is, symmetric in the identical subsystems and no permuting 
symmetry between charm and bottom quarks. Also, the orbital parts are 
symmetric for the $S-$wave ground states. Then, the color-spin wave 
functions should be antisymmetric in the charm or bottom subsystem. All 
the possible configurations for the $S-$wave fully-heavy hexaquark 
systems are listed in Table~\ref{configuration}. The explicit forms for 
color-spin wave functions can be obtained with the help of 
Clebsch-Gordan coefficients of groups $SU(N)$ and $S_N$ after 
tedious calculations~\cite{deSwart:1963pdg,Kaeding:1995vq,
Stancu:1991rc,Park:2015nha}. With the color-spin-flavor wave 
functions, one can calculate the matrix elements of operators 
${\vlab}_i\cdot{\vlab}_j$ and 
${\vlab}_i\cdot{\vlab}_j{\vsig}_i\cdot{\vsig}_j$ in the Hamiltonian. 
Here, we present the final elements for all configurations in 
Table~\ref{sfc} for reference. \\    

\begin{table}[!htbp]
	\begin{center}
		\caption{ \label{configuration} All possible configurations for $S-$wave fully-heavy hexaquark systems. The subscripts and superscripts stand for the spin and color types, respectively. The braces $\{ ~\}$ are adopted for the subsystems with symmetric flavor wave functions.}
		\begin{tabular*}{8.6cm}{@{\extracolsep{\fill}}p{1.5cm}<{\centering}p{1.5cm}<{\centering}p{2.5cm}<{\centering}p{2.5cm}<{\centering}}
			\hline\hline
			System & $J^P$         &  \multicolumn{2}{c}{Configuration} \\\hline
			$cccccc$ & $0^+$       &  $|\{cccccc\}_0^1\rangle_0^1$
			&   $\cdots$      \\
			$cccccb$ & $0^+$       &  $|\{ccccc\}_{1/2}^{\bar 3} b_{1/2}^3\rangle_0^1$
			&   $\cdots$      \\
			& $1^+$       &  $|\{ccccc\}_{1/2}^{\bar 3} b_{1/2}^3\rangle_1^1$
			&   $\cdots$      \\
			$ccccbb$ & $0^+$       &  $|\{cccc\}_{1}^{3}\{bb\}_{1}^{\bar 3}\rangle_0^1$
			&   $|\{cccc\}_{0}^{\bar 6}\{bb\}_{0}^{6}\rangle_0^1$      \\
			& $1^+$       &  $|\{cccc\}_{1}^{3}\{bb\}_{1}^{\bar 3}\rangle_1^1$
			&   $\cdots$      \\
			& $2^+$       &  $|\{cccc\}_{1}^{3}\{bb\}_{1}^{\bar 3}\rangle_2^1$
			&   $\cdots$      \\
			$cccbbb$ & $0^+$       & $|\{ccc\}_{3/2}^{1}\{bbb\}_{3/2}^{1}\rangle_0^1$
			&   $|\{ccc\}_{1/2}^{8}\{bbb\}_{1/2}^{8}\rangle_0^1$      \\
			& $1^+$       & $|\{ccc\}_{3/2}^{1}\{bbb\}_{3/2}^{1}\rangle_1^1$
			&   $|\{ccc\}_{1/2}^{8}\{bbb\}_{1/2}^{8}\rangle_1^1$      \\
			& $2^+$       & $|\{ccc\}_{3/2}^{1}\{bbb\}_{3/2}^{1}\rangle_2^1$
			&   $\cdots$      \\
			& $3^+$       & $|\{ccc\}_{3/2}^{1}\{bbb\}_{3/2}^{1}\rangle_3^1$
			&   $\cdots$      \\
			$bbbbcc$ & $0^+$       &  $|\{bbbb\}_{1}^{3}\{cc\}_{1}^{\bar 3}\rangle_0^1$
			&   $|\{bbbb\}_{0}^{\bar 6}\{cc\}_{0}^{6}\rangle_0^1$      \\
			& $1^+$       &  $|\{bbbb\}_{1}^{3}\{cc\}_{1}^{\bar 3}\rangle_1^1$
			&   $\cdots$      \\
			& $2^+$       &  $|\{bbbb\}_{1}^{3}\{cc\}_{1}^{\bar 3}\rangle_2^1$
			&   $\cdots$      \\
			$bbbbbc$ & $0^+$       &  $|\{bbbbb\}_{1/2}^{\bar 3} c_{1/2}^3\rangle_0^1$
			&   $\cdots$      \\
			& $1^+$       &  $|\{bbbbb\}_{1/2}^{\bar 3} c_{1/2}^3\rangle_1^1$
			&   $\cdots$      \\
			$bbbbbb$ & $0^+$       &  $|\{bbbbbb\}_0^1\rangle_0^1$
			&   $\cdots$     \\\hline
		\end{tabular*}
	\end{center}
\end{table}
 
 \begin{table*}[!htbp]
 	\begin{center}
 		\caption{ \label{sfc} Color-spin matrix elements for the $cccccc$, $cccccb$, $ccccbb$, and $cccbbb$ systems. The color-spin matrix elements for the $bbbbbb$, $bbbbbc$, and $bbbbcc$ systems are same as that of $cccccc$, $cccccb$, and $ccccbb$ systems by interconverting the charm and bottom quarks, respectively.}
 			\begin{tabular*}{18cm}{@{\extracolsep{\fill}}p{1cm}<{\centering}p{5cm}<{\centering}p{1cm}<{\centering}p{1cm}<{\centering}p{1cm}<{\centering}p{1cm}<{\centering}p{1cm}<{\centering}p{1cm}<{\centering}}
 			\hline\hline
 		  \multirow {2}{*}{System} &  \multirow {2}{*}{$\langle \hat O \rangle$}  & \multicolumn{3}{c}{${\vlab}_i\cdot{\vlab}_j$}  & \multicolumn{3}{c}{${\vlab}_i\cdot{\vlab}_j {\vsig}_i\cdot{\vsig}_j$} \\\cline{3-5}\cline{6-8}
 		  & & $cc$     & $cb$  & $bb$   & $cc$     & $cb$  & $bb$ \\\hline
 			
 			$cccccc$    &  $^1_0 \langle \{cccccc\}_0^1| \hat O  |\{cccccc\}_0^1\rangle_0^1$   & $-16/15$ & $\cdots$  & $\cdots$  & $-16/5$  &  $\cdots$   &  $\cdots$  \\
 			
 		    $cccccb$    &  $^1_0 \langle \{ccccc\}_{1/2}^{\bar 3} b_{1/2}^3| \hat O |\{ccccc\}_{1/2}^{\bar 3} b_{1/2}^3\rangle_0^1$   & $-16/15$ &  $-16/15$ & $\cdots$  & $-16/5$  &  $-16/5$  &  $\cdots$  \\ 
 		   &  $^1_1 \langle \{ccccc\}_{1/2}^{\bar 3} b_{1/2}^3| \hat O |\{ccccc\}_{1/2}^{\bar 3} b_{1/2}^3\rangle_1^1$   & $-16/15$ &  $-16/15$ & $\cdots$  & $-16/5$  &  $16/15$  &  $\cdots$  \\
 			
 			$ccccbb$    &  $_0^1 \langle \{cccc\}_{1}^{3}\{bb\}_{1}^{\bar 3}| \hat O |\{cccc\}_{1}^{3}\{bb\}_{1}^{\bar 3}\rangle_0^1$   & $-4/3$ &  $-2/3$ & $-8/3$  & $-28/9$  &  $-4/3$  &  $-8/3$  \\ 
 			&  $ _0^1 \langle \{cccc\}_{0}^{\bar 6}\{bb\}_{0}^{6}| \hat O  |\{cccc\}_{0}^{\bar 6}\{bb\}_{0}^{6}\rangle_0^1$   & $-2/3$ &  $-5/3$ & $4/3$  & $-10/3$  &  $0$  &  $-4$  \\ 
 			&  $_0^1 \langle \{cccc\}_{1}^{3}\{bb\}_{1}^{\bar 3}| \hat O  |\{cccc\}_{0}^{\bar 6}\{bb\}_{0}^{6}\rangle_0^1$   & $0$ &  $0$ & $0$  & $0$  &  $\sqrt{6}$  &  $0$  \\ 
 			
 		    &  $_1^1 \langle \{cccc\}_{1}^{3}\{bb\}_{1}^{\bar 3}| \hat O |\{cccc\}_{1}^{3}\{bb\}_{1}^{\bar 3}\rangle_1^1$   & $-4/3$ &  $-2/3$ & $-8/3$  & $-28/9$  &  $-2/3$  &  $-8/3$  \\ 
 		   &  $_2^1 \langle \{cccc\}_{1}^{3}\{bb\}_{1}^{\bar 3}| \hat O |\{cccc\}_{1}^{3}\{bb\}_{1}^{\bar 3}\rangle_2^1$    & $-4/3$ &  $-2/3$ & $-8/3$  & $-28/9$  &  $2/3$  &  $-8/3$  \\ 
 			
 	  	$cccbbb$    &  $_0^1 \langle \{ccc\}_{3/2}^{1}\{bbb\}_{3/2}^{1}| \hat O |\{ccc\}_{3/2}^{1}\{bbb\}_{3/2}^{1}\rangle_0^1$   & $-8/3$ &  $0$ & $-8/3$  & $-8/3$  &  $0$  &  $-8/3$  \\ 
 	  	&  $_0^1 \langle \{ccc\}_{1/2}^{8}\{bbb\}_{1/2}^{8}| \hat O |\{ccc\}_{1/2}^{8}\{bbb\}_{1/2}^{8}\rangle_0^1$   & $-2/3$ &  $-4/3$ & $-2/3$  & $-10/3$  &  $-20/9$  &  $-10/3$  \\ 
 	  	& $_0^1 \langle \{ccc\}_{3/2}^{1}\{bbb\}_{3/2}^{1}| \hat O |\{ccc\}_{1/2}^{8}\{bbb\}_{1/2}^{8}\rangle_0^1$   & $0$ &  $0$ & $0$  & $0$  &  $-16/9$  &  $0$  \\ 
 	  	
 	  	 &  $_1^1 \langle \{ccc\}_{3/2}^{1}\{bbb\}_{3/2}^{1}| \hat O |\{ccc\}_{3/2}^{1}\{bbb\}_{3/2}^{1}\rangle_1^1$   & $-8/3$ &  $0$ & $-8/3$  & $-8/3$  &  $0$  &  $-8/3$  \\ 
 	  	&  $_1^1 \langle \{ccc\}_{1/2}^{8}\{bbb\}_{1/2}^{8}| \hat O |\{ccc\}_{1/2}^{8}\{bbb\}_{1/2}^{8}\rangle_1^1$   & $-2/3$ &  $-4/3$ & $-2/3$  & $-10/3$  &  $20/27$  &  $-10/3$  \\ 
 	  	& $_1^1 \langle \{ccc\}_{3/2}^{1}\{bbb\}_{3/2}^{1}| \hat O |\{ccc\}_{1/2}^{8}\{bbb\}_{1/2}^{8}\rangle_1^1$   & $0$ &  $0$ & $0$  & $0$  &  $-16\sqrt{5}/27$  &  $0$  \\ 
 	  	
 	  	&  $_2^1 \langle \{ccc\}_{3/2}^{1}\{bbb\}_{3/2}^{1}| \hat O |\{ccc\}_{3/2}^{1}\{bbb\}_{3/2}^{1}\rangle_2^1$   & $-8/3$ &  $0$ & $-8/3$  & $-8/3$  &  $0$  &  $-8/3$  \\ 
 	  	
 	  	&  $_3^1 \langle \{ccc\}_{3/2}^{1}\{bbb\}_{3/2}^{1}| \hat O |\{ccc\}_{3/2}^{1}\{bbb\}_{3/2}^{1}\rangle_3^1$   & $-8/3$ &  $0$ & $-8/3$  & $-8/3$  &  $0$  &  $-8/3$  \\ 
 			
 			\hline\hline
 		\end{tabular*}
 	\end{center}
 \end{table*}
The trail orbital wave function for a hexaquark state in the coordinate space can be expanded by a series of Gaussian functions,
\begin{eqnarray}\label{gsfa}
	\psi({\mathbf{r_1},\mathbf{r_2},\mathbf{r_3},\mathbf{r_4},\mathbf{r_5},\mathbf{r_6}})&=&\sum^n_{\ell} \mathcal{C}_{\ell}\prod_{i=1}^6\left(\frac{m_i\omega_{\ell}}{\pi}\right)^{3/4}\exp\left[-\frac{m_i\omega_l}{2}r^2_i\right]\nonumber\\
	&\equiv&\sum^n_{\ell} \mathcal{C}_{\ell}\prod_{i=1}^6 \phi(\omega_\ell,\mathbf{r}_i),
\end{eqnarray}
which is usually adopted to study the compact multiquark 
systems~\cite{Zhang:2007mu,Zhang:2005jz,Liu:2019zuc}. Some useful 
overlaps for Gaussian functions are shown as follows

\begin{equation}\label{T Jacobi}
	\left\langle\prod_{j=1}^6 \phi(\omega_\ell,\mathbf{r}_j)\left| \sum_{i=1}^6T_i-T_G\right| \prod_{k=1}^6 \phi(\omega_\ell,\mathbf{r}_k)\right\rangle
	=3840\frac{(\omega_\ell\omega_{\ell'})^{11/2}}{(\omega_\ell+\omega_{\ell'})^{10}}.
\end{equation}

\begin{equation}\label{r matrix element}
	\left\langle\psi(\omega_\ell,\mathbf r_{i}, \mathbf r_{j})\left|\frac{1}{r_{ij}}\right|\psi(\omega_{\ell'},\mathbf r_{i}, \mathbf r_{j})\right\rangle=2\sqrt{\frac{m_{ij}}{\pi}}\frac{(\omega_\ell\omega_{\ell'})^{3/2}}{(\frac{\omega_\ell+\omega_{\ell'}}{2})^{5/2}},
\end{equation}
\begin{equation}\label{sigp matrix element}
	\left\langle\psi(\omega_{\ell},\mathbf r_{i}, \mathbf r_{j})\left|e^{-\sigma^2_{ij}r_{ij}^2}\right|\psi(\omega_{\ell'},\mathbf r_{i}, \mathbf r_{j})\right\rangle
	=\left(\frac{m_{ij}(\frac{2\omega_\ell\omega_{\ell'}}{\omega_\ell+\omega_{\ell'}})}{m_{ij}\frac{\omega_\ell+\omega_{\ell'}}{2}+\sigma_{ij}^2}\right)^{\frac{3}{2}},
\end{equation}
\begin{equation}\label{sigp matrix element}
	\left\langle\psi(\omega_\ell,\mathbf r_{i}, \mathbf r_{j})\left| r_{ij}\right|\psi(\omega_{\ell'},\mathbf r_{i}, \mathbf r_{j})\right\rangle=2\sqrt{\frac{1}{\pi m_{ij}}}\frac{(\omega_\ell\omega_{\ell'})^{3/2}}{(\frac{\omega_\ell+\omega_{\ell'}}{2})^{7/2}},
\end{equation}
where $\psi(\omega_i,\mathbf r_{i}, \mathbf r_{j})\equiv \phi(\omega_\ell,\mathbf{r}_i)\phi(\omega_\ell,\mathbf{r}_j)$, $m_{ij}=m_im_j/(m_i+m_j)$.\\

\quad
With the full wave functions and all the matrix elements involved in the 
Hamiltonian, the masses without a mixing mechanism can therefore be 
calculated by solving the generalized eigenvalue problem 
straightforwardly
\begin{equation}
	\sum_{\ell'=1}^{n}(H_{\ell \ell'}-EN_{\ell \ell'})C_{\ell'}=0,~~~~ (\ell=1, 2, ... , n),
\end{equation}
where the $H_{\ell \ell'}$ are the matrix elements of the total 
Hamiltonian, $N_{\ell \ell'}$ are the overlap matrix elements of the 
Gaussian functions arising from their nonorthogonality, $E$ stands for 
the mass, and $\mathcal{C}_{\ell'}$ is the eigenvector corresponding to 
the coefficients of the orbital wave function for a fully-heavy 
hexaquark. Moreover, for a given hexaquark system, different 
configurations with same $J^P$ can mix with each other in principle. 
The mixing effects are taken into account and discussed in present 
calculations, and then the final mass spectra and wave functions are 
obtained by diagonalizing the mass matrix of these configurations.

\section{Mass spectra and discussions}{\label{results}}

\begin{figure*}[!htbp]
	\centering
	\includegraphics[scale=0.75]{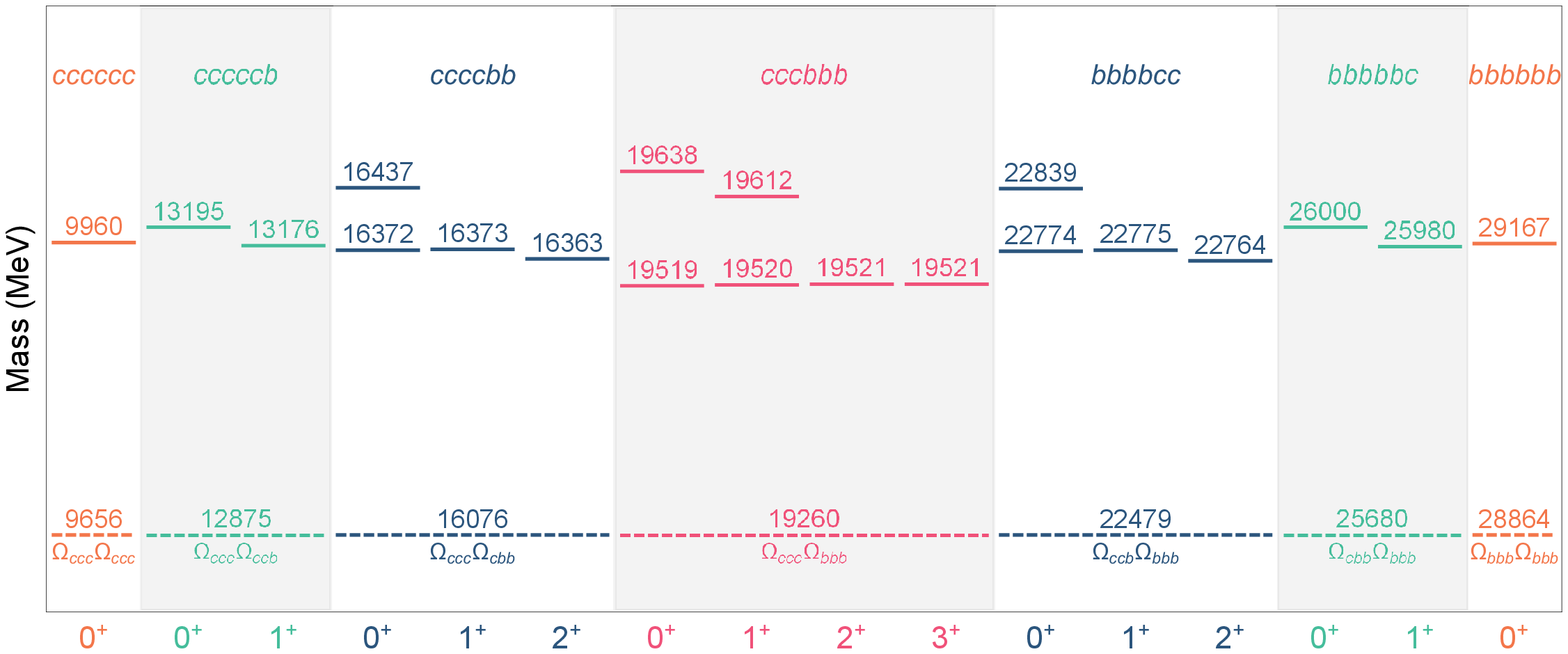}
	\vspace{0.0cm} \caption{Mass spectra for the fully-heavy hexaquarks. The solid lines stand for the predicted masses of different systems, and dashed lines are their corresponding thresholds.}
	\label{mass}
\end{figure*}

\quad In this work, we adopt the same variational parameters 
$\omega_\ell$ and $n$ as previous work~\cite{Liu:2019zuc} to solve the  
generalized eigenvalue problem numerically. With these parameters, 
we can obtain stable mass spectra of fully-heavy tetraquarks, and 
also fully-heavy hexaquarks here. Certainly, these variational 
parameters are introduced just for a numerical calculation and 
do not affect the final results of constituent quark model when the 
numerical procedure is convergent and stable 
enough~\cite{Hiyama:2003cu}. The predicted masses for $S-$ wave 
fully-heavy hexaquarks are listed in Table~\ref{mass1} and 
Figure~\ref{mass}.  \\

\begin{table*}[htp]
	\begin{center}
		\caption{\label{mass1} The predicted masses for $S-$wave fully-heavy hexaquarks.}
		\begin{tabular*}{18cm}{@{\extracolsep{\fill}}p{1cm}<{\centering}p{1cm}<{\centering}p{1.7cm}<{\centering}p{3cm}<{\centering}p{1.8cm}<{\centering}p{3cm}<{\centering}p{2.3cm}<{\centering}}
			\hline\hline
			System &	$J^{P}$  & Configuration                                             & $\langle H\rangle$ (MeV) & Mass (MeV)  & Eigenvector & Threshold (MeV) \\\hline
			
			$cccccc$ & $0^+$       &  $|\{cccccc\}_0^1\rangle_0^1$   & 9960 & 9960 & 1 & 9656\\
			
			$cccccb$ & $0^+$       &  $|\{ccccc\}_{1/2}^{\bar 3} b_{1/2}^3\rangle_0^1$   & 13195 & 13195 & 1  & 12875 \\
			& $1^+$       &  $|\{ccccc\}_{1/2}^{\bar 3} b_{1/2}^3\rangle_1^1$  & 13176 & 13176 & 1  & 12875  \\
			
			$ccccbb$ & $0^+$       &  \multirow{2}{*}{$\begin{bmatrix}  |\{cccc\}_{1}^{3}\{bb\}_{1}^{\bar 3}\rangle_0^1 \\ |\{cccc\}_{0}^{\bar 6}\{bb\}_{0}^{6}\rangle_0^1 \end{bmatrix}$} 
			& \multirow{2}{*}{$\begin{pmatrix}16378&-18 \\-18&16431 \end{pmatrix}$}
			& \multirow{2}{*}{$\begin{bmatrix}16372 \\16437 \end{bmatrix}$}  & \multirow{2}{*}{$\begin{bmatrix}(-0.956,0.295)\\(0.295, 0.956) \end{bmatrix}$} & 16076 \\
			& & & & & & 16076\\
			& $1^+$       &  $|\{cccc\}_{1}^{3}\{bb\}_{1}^{\bar 3}\rangle_1^1$  & 16373 & 16373 & 1 & 16076  \\
			& $2^+$       &  $|\{cccc\}_{1}^{3}\{bb\}_{1}^{\bar 3}\rangle_2^1$  & 16363 & 16363 & 1 & 16076  \\
			
			$cccbbb$ & $0^+$       &  \multirow{2}{*}{$\begin{bmatrix}  |\{ccc\}_{3/2}^{1}\{bbb\}_{3/2}^{1}\rangle_0^1 \\ |\{ccc\}_{1/2}^{8}\{bbb\}_{1/2}^{8}\rangle_0^1 \end{bmatrix}$} 
			& \multirow{2}{*}{$\begin{pmatrix}19521&-15 \\-15&19636\end{pmatrix}$}
			& \multirow{2}{*}{$\begin{bmatrix}19519 \\19638 \end{bmatrix}$}  & \multirow{2}{*}{$\begin{bmatrix}(-0.992,0.130)\\(0.130, 0.992) \end{bmatrix}$} & 19260\\
			& & & & & & 19260 \\
			& $1^+$       &  \multirow{2}{*}{$\begin{bmatrix}  |\{ccc\}_{3/2}^{1}\{bbb\}_{3/2}^{1}\rangle_1^1 \\ |\{ccc\}_{1/2}^{8}\{bbb\}_{1/2}^{8}\rangle_1^1 \end{bmatrix}$} 
			& \multirow{2}{*}{$\begin{pmatrix}19521&-12 \\-12&19610\end{pmatrix}$}
			& \multirow{2}{*}{$\begin{bmatrix}19520 \\19612 \end{bmatrix}$}  & \multirow{2}{*}{$\begin{bmatrix}(-0.992,0.127)\\(0.127, 0.992) \end{bmatrix}$} & 19260\\
			& & & & & & 19260 \\
			& $2^+$       & $|\{ccc\}_{3/2}^{1}\{bbb\}_{3/2}^{1}\rangle_2^1$ & 19521 & 19521 & 1 & 19260  \\
			& $3^+$      	& $|\{ccc\}_{3/2}^{1}\{bbb\}_{3/2}^{1}\rangle_3^1$ & 19521 & 19521 & 1  & 19260 \\	
			
			$bbbbcc$ & $0^+$       &  \multirow{2}{*}{$\begin{bmatrix}  |\{bbbb\}_{1}^{3}\{cc\}_{1}^{\bar 3}\rangle_0^1 \\ |\{bbbb\}_{0}^{\bar 6}\{cc\}_{0}^{6}\rangle_0^1 \end{bmatrix}$} 
			& \multirow{2}{*}{$\begin{pmatrix}22780&-19 \\-19&22833 \end{pmatrix}$}
			& \multirow{2}{*}{$\begin{bmatrix}22774 \\22839 \end{bmatrix}$}  & \multirow{2}{*}{$\begin{bmatrix}(-0.953,0.302)\\(0.302, 0.953) \end{bmatrix}$} & 22479\\
			& & & & & & 22479\\
			& $1^+$       &  $|\{bbbb\}_{1}^{3}\{cc\}_{1}^{\bar 3}\rangle_1^1$ & 22775 & 22775 & 1 & 22479  \\
			& $2^+$       &  $|\{bbbb\}_{1}^{3}\{cc\}_{1}^{\bar 3}\rangle_2^1$  & 22764 & 22764 & 1 & 22479  \\
			
			$bbbbbc$ & $0^+$       &  $|\{ccccc\}_{1/2}^{\bar 3} b_{1/2}^3\rangle_0^1$   & 26000 & 26000 & 1  & 25680 \\
			& $1^+$       &  $|\{ccccc\}_{1/2}^{\bar 3} b_{1/2}^3\rangle_1^1$  & 25980 & 25980 & 1 & 25680  \\
			
			$bbbbbb$ & $0^+$       &  $|\{bbbbbb\}_0^1\rangle_0^1$   & 29167 & 29167 & 1 & 28864 \\
			
			\hline\hline
		\end{tabular*}
	\end{center}
\end{table*}

\quad It can be seen that all of the fully-heavy hexaquarks  are 
predicted to lie above the corresponding thresholds. These results 
are consistent with other quark model 
calculations~\cite{Richard:2020zxb,Alcaraz-Pelegrina:2022fsi,
Weng:2022ohh} but contradict with conclusions of lattice 
QCD~\cite{Junnarkar:2019equ,Lyu:2021qsh,Mathur:2022nez}, one-boson 
exchange model~\cite{Liu:2021pdu}, and QCD sum rule~\cite{Wang:2022jvk}. 
Instead of the absolute masses, whether above or below the threshold is 
just a qualitative conclusion, which can not be simply boiled down 
to the uncertainties of models and numerical calculations. Our 
results together with previous quark model works indicate that the 
potential with Coulomb potential, linear confining potential, and 
spin-spin interactions can hardly give a stable compact fully-heavy 
hexaquark. Probably, more interactions or mechanism are included 
in lattice QCD approach, which lead to this difference. More 
theoretical works and future experimental data will help us to 
resolve this puzzle.\\ 

\quad With the obtained wave functions, we can also estimate the 
expectations of $\langle \boldsymbol r_{ij}^2  \rangle^{1/2}$ for $cc$, 
$cb$, and $bb$ subsystems. We find that these expectations are similar 
in different configurations, and are about $0.53 \sim 0.55$, $0.43 \sim 
0.44$, and $0.29 \sim 0.30$ fm for $cc$, $cb$, and $bb$ subsystems, 
respectively.  These relatively small distances show typical features of 
the compact multiquarks, which are significantly different from the 
loosely bound molecular picture. Moreover, for $ccccbb$, $cccbbb$, and 
$bbbbcc$ systems, some configurations with same $J^P$ can mix with each 
other. The mixing effects arise from the spin-spin interaction, and 
then are highly suppressed by the masses of heavy quarks. Hence, it 
is inevitable that these mixtures should be extremely small, and our 
numerical results confirm this expectation. \\

\quad An interesting discovery is the inverted mass hierarchy for 
several fully-heavy hexaquarks. Usually, one expect that the 
mass of higher $J$ state should be larger than that of the lower $J$ 
one for a given spin multiplicity, which is also confirmed by the 
spectroscopy of conventional hadrons both theoretically and \
experimentally. In the 
literature~\cite{Isgur:1998kr,WooLee:2006kdh,Liang:2019aag,Luo:2019qkm}, 
the authors discussed the possibilities of inverted mass hierarchy
for conventional hadrons by considering specific dynamical mechanism, 
however, these predictions are not verified by experiments until now. In 
the fully-heavy hexaquarks, the lightest states for the $cccccb$, 
$ccccbb$, $bbbbbc$, and $bbbbcc$ systems have the highest total angular 
momentum. Moreover, for the pure configurations, when the total angular 
momentum increases, the masses decrease. This inverted mass hierarchy 
is due in the symmetry constraint, where the color-spin wave functions 
are limited in specific forms and jointly act on the fine splittings. 
Especially, the four pure configurations 
$|\{ccc\}_{3/2}^{1}\{bbb\}_{3/2}^{1}\rangle_S^1$ with $S=0, 1, 2, 3$ 
have the same mass because no residual interaction is left between the 
$ccc$ and $bbb$ subsystems. It can be seen that the exotic states 
have more complicated color structures than that of conventional 
hadrons, and thus can perform more particular spectrum. We hope that 
the future experiments can test this inverted mass hierarchy of 
fully-heavy hexaquarks. \\       

\begin{figure}[!htbp]
	\centering
	\includegraphics[scale=0.58]{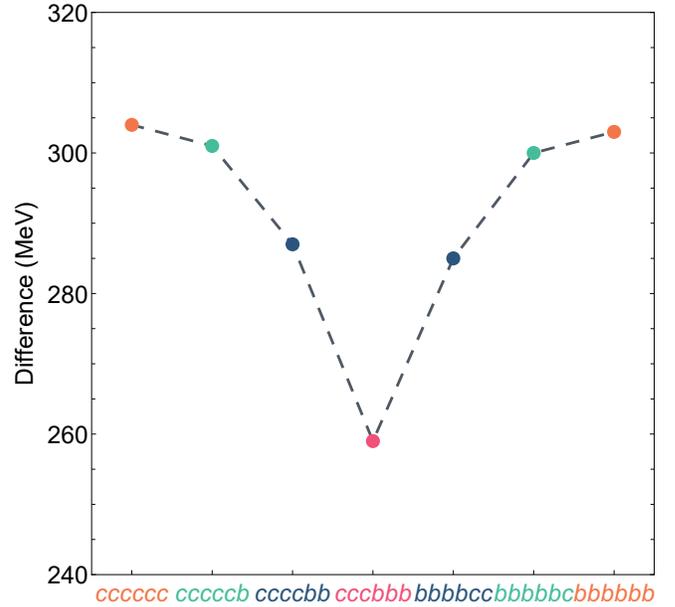}
	\vspace{0.0cm} \caption{Mass differences between the lowest hexaquarks and their corresponding thresholds for different systems. }
	\label{dif}
\end{figure}

\quad We can also investigate the mass differences between hexaquark 
states and their corresponding thresholds versus different systems. 
The mass differences between the lowest hexaquarks and
thresholds for different systems are plotted in Figure~\ref{dif}. It can 
be found that the $cccbbb$ system has lowest mass difference while the 
$cccccc$ and $bbbbbb$ systems have larger mass differences. 
Compared with the $cccccc$ and $bbbbbb$ systems, the $cccbbb$ 
system seems to be more asymmetric. Also, if one splits a 
fully-heavy hexaquark into two three-quark subsystems, it is easy 
to see that the $cccbbb$ system has the largest mass ratio between 
these two subsystems. Actually, in previous study on 
tetraquarks~\cite{Lu:2020rog}, we also found that the systems 
with larger mass ratios tend to be stable. Our present results 
indicate that the fully-heavy hexaquarks with lower symmetry and 
larger mass ratios are more likely to stabilize.  If one keeps 
increasing the mass ratios, the fully-heavy hexaquarks will 
become heavy-light systems. Thus, it is reasonable to speculate 
that there may be some stable heavy-light hexaquarks below the 
corresponding thresholds. More precise calculations for the 
heavy-light hexaquarks within our framework are needed to verify 
or deny this conjecture. 

Furthermore, it should be emphasized that, even if a exotic state 
is not bound, it may also subsist as a resonance with finite decay 
width and be observed by future experiments. These fully-heavy 
hexaquarks can decay into the two fully-heavy baryons by 
fall-apart mechanism. The precise decay widths depend on the 
transition operator, wave functions of initial and final states, 
and phase space, which are all model dependent. Without any 
experimental data as a benchmark, it is hard to estimate the total 
widths of these resonance theoretically at present.\\

\section{Summary}{\label{Summary}}

\quad In this paper, we adopt the norelativistic constituent quark 
model to investigate the S-wave fully-heavy hexaquarks systematically. 
The mass spectra are obtained by solving the Hamiltonian including the 
Coulomb potential, confining potential, and spin-spin interactions. 
All of the fully-heavy hexaquarks are predicted to lie above the 
corresponding baryon-baryon thresholds, and thus no stable binding 
one exists. Then, these fully-heavy hexaquarks can subsist as 
resonances and may easily decay into the fully-heavy baryons through 
the fall-apart mechanism. Our present results are consistent with other 
quark model calculations but different with the conclusions of lattice 
QCD, QCD sum rule, and one-boson exchange model. More theoretical 
efforts and future experimental data will help us to resolve this 
puzzle.\\

\quad With the obtained wave functions, we estimate the 
expectations of $\langle \boldsymbol r_{ij}^2  \rangle^{1/2}$ for $cc$, 
$cb$, and $bb$ subsystems, and find that they have typical sizes of 
compact multiquarks. Also, we find that some fully-heavy hexaquark have 
inverted mass hierarchy, that is, the masses of pure configurations 
decrease when the total angular momentum increases. Moreover, the mass 
differences between hexaquark states and their corresponding thresholds 
for different systems are discussed, which can provide clues for us to 
hunting for stable hexaquarks. Thus, we speculate that there may be 
some stable heavy-light hexaquarks below the corresponding thresholds. 
We hope our present predictions can provide helpful information for 
future experimental searches.

\bigskip
\noindent
\begin{center}
	{\bf ACKNOWLEDGEMENTS}\\
\end{center}

We would like to thank Ming-Sheng Liu for valuable discussions. Qi-Fang 
L\"u thanks Atsushi Hosaka for his warm hospitality in Osaka University. 
This work is supported by the National Natural Science Foundation of 
China under Grants No.11705056, No. 12175037, No.11947224, No.11475192, 
No.11975245, and No.U1832173. This work is also supported by the Key 
Project of Hunan Provincial Education Department under 
Grant No. 21A0039, the State Scholarship Fund of China 
Scholarship Council under Grant No. 202006725011, the Sino-German 
CRC 110 “Symmetries and the Emergence of Structure in QCD” project by 
NSFC under the Grant No. 12070131001, the Key Research Program of 
Frontier Sciences, CAS, under the Grant No. Y7292610K1,
and the National Key Research and Development Program of China under 
Contracts No. 2020YFA0406300.

\end{document}